\documentclass[aps,twocolumn,showpacs,superscriptaddress,preprintnumbers,amsmath,amssymb]{revtex4}

\usepackage{graphicx}
\usepackage{amsmath,amsthm,amsfonts,amscd}

\begin{document}

\title{Collisionless filamentation, filament merger and heating of
low-density relativistic electron beam propagating through a background
plasma}
\author{Vladimir Khudik}
\affiliation{Department of Physics and Institute for Fusion Studies, The University of Texas at Austin, TX 78712, U.S.A.}
\author{Igor Kaganovich}
\affiliation{Plasma Physics Laboratory, Princeton University, Princeton, New Jersey 08543, USA}
\author{Gennady Shvets}
\affiliation{Department of Physics and Institute for Fusion Studies, The University of Texas at Austin, TX 78712, U.S.A.}
\email{vvv@aaa.edu}

\date{\today }

\begin{abstract}
A cold electron beam propagating through a background plasma is subject to
filamentation process due to the Weibel instability. If the initial beam
radius is large compared with the electron skin depth and the beam density
is much smaller than the background plasma density, multiple filaments merge
many times. Because of this non-adiabatic process, the beam perpendicular
energy of initially cold beam grows until all filaments coalesce into one
pinched beam with the beam radius much smaller than initial radius and
smaller than the electron skin depth. It was shown through particle-in-cell
simulations that a significant fraction of the beam is not pinched  by the magnetic forces of the pinched beam and fills most of the
plasma region. The resulting electron beam energy distribution in the perpendicular
direction is close to a Maxwellian for the bulk electrons. However, there
are significant departures from a Maxwellian for low and high perpendicular
energy (deeply trapped and untrapped electrons). An analytical model is
developed describing the density profile of the resulting pinched beam and
large low-density halo around it. Based on this analytical model, a
calculation of the energy transfer from the beam longitudinal kinetic energy
to the transverse beam kinetic energy, the self-magnetic field, and the
plasma electrons is performed. Results of analytical theory agree well with
the particle-in-cell simulations results.
\end{abstract}


\maketitle

\section{Introduction}

\bigskip The Weibel instability (WI)~\cite{Weibel,Fried,Morse,David,Lampe}
of plasmas with anisotropic velocity distribution is one of the most basic
and long-studied collective plasma processes. For example, propagation of an
electron beam through the background plasma is subject to strong WI~\cite%
{Fried}. There has been a significant revival in theoretical studies of the
WI because it is viewed as highly relevant to at least two area of science:
astrophysics of gamma-ray bursts and their afterglows~\cite%
{Medv,Medv1,Gruz,Spitk,Mor1,Milos} and the Fast Ignitor~\cite{Tabak}
scenario of the inertial confinement fusion (ICF). Specifically, generation
of the upstream magnetic field during the GRB aftershocks is considered
necessary for explaining emission spectra of the afterglows as well as for
generating and sustaining collisionless shocks responsible for particle
acceleration during GRBs. Collisionless Weibel Instability is the likeliest
mechanism~\cite{Medv,Medv1,Gruz,Spitk,Mor1,Milos} for producing such
magnetic fields. The WI is likely to play an important role in the Fast
Ignitor scenario~\cite{Tabak} because it can result in the collective energy
loss of a relativistic electron beam in both coronal and core plasma regions 
\cite{Tabak, Honrub,Taguchi,Mor2,Book,Us,malkin_02,key_pop05,honda_pukhov}.
Because the relativistic electron beam has to travel through an enormous
density gradient (varying from $10^{22}$cm$^{-3}$ near the critical surface
where the beam is produced to $10^{26}$cm$^{-3}$ in the dense core), both
collisionless and collisional WI manifest themselves along the beam's path.

The dynamics and energetics of the nonlinear saturation and long-term
behavior of the WI are important for both laboratory and astrophysical
plasmas. For example, collisionless shock dynamics depends on the long-term
evolution of the magnetic field energy. Specifically, it is not clear
whether the long-term magnetic fields generated during the coalescence of
current filaments remain finite~\cite{Mor2} or decay with time~\cite{Gruz}
(and if they do, according to what physical mechanism). Numerous numerical
simulations~\cite{Morse,Silva_ApJ03} demonstrated that magnetic field energy
grows during the earlier stages of the WI and starts decaying during the
later (strongly nonlinear) stage. The reason for this decay has never been
fully understood. One decay mechanism based on the merger of filaments
bearing super-Alfvenic current ($I>I_{A}=\gamma \beta mc^{3}/e$~\cite%
{DavBook}, where $-e$ and $m$ are the electron charge and mass,
respectively, and $c$ is the speed of light in vacuum) during the late stage
of the WI has been recently identified~\cite{shvets_prl08}. In the review 
\cite{shvets_pop09} we presented a detailed analysis of the high-current
filaments' current and density profiles and provide qualitative and
quantitative explanation of the energetics of their merger in the limit ($%
I>I_{A}$). In that case the filaments carry super Alfenic current and their
perpendicular energy distribution is closely described by a RH distribution
that is $\delta -$ function of perpendicular kinetic energy, or KV
distribution as it is called in accelerator physics. For such a distribution
functions the density profile in the filament is flat and the radial
electric field vanishes. Particle-in-cell simulation of the nonlinear stages
of the Weibel instability showed significant ion acceleration in the radial
electric field. \textbf{Therefore, it is very important to investigate
departure from RH or KV distribution during nonlinear stages of the Weibel
instability, which ultimately determines in acceleration in the micro field
of the filammetns. }

To that end we investigated propagation of the relativistic electron beam
through a background plasma and development of the self-electric and
magnetic field. Of particular interest is a case when the beam transverse
size is much larger than the electron skin depth by a factor of ten and
more, $r_{b}>10\delta _{p}$. The filament resulting from the Weibel
instability are typically of the size of the electron skin depth, $\delta
_{p}$. Here,$\ \delta _{p}=c/\omega _{p}$, $\omega _{p}=\sqrt{4\pi e^{2}np/m%
}$ is the electron plasma frequency, and $n_{p}$ is the uniform background
plasma density. Therefore in the limit $r_{b}>10\delta _{p}$ many, 
$>100$ filaments are formed and then merge multiple times. In the
process of merger there is an effective exchange in the perpendicular energy
between beam particles and plasma electrons. If the beam density approaches
the background plasma density, the beam electrons expel the plasma electrons
and the beam electron space charge is neutralized by the background ions. In
this case the further pinching of the beam is limited by the background
plasma density. To avoid this limitation, we studied the very low density
beam with the density a factor of 1000 less than the background plasma
density. This insured that as beam filaments merge and the beam density
dramatically increases the beam density still remains small compared with
the plasma density. In this limiting case, common PIC codes are not
efficient for the description of the plasma electron due to the large
numerical noise compared to the beam density. However, we can utilize a
semi-analytic approach for description of the plasma electrons. Because the
beam evolution occur on a time scale much smaller than the electron plasma
period, the electron background plasma adiabatically modifies to the beam
current profile via the return current, see Ref. \cite{shvets_pop09} for
details. We also assume charge neutrality of the system consisting of the
beam electrons, ambient plasma electrons, and ambient plasma ions~\cite{Us}.
These simulations do not resolve the motion of the ambient plasma electrons
and, therefore, can take computational time steps $\Delta t>1/\omega_p $.
After exclusion of fast motions of the plasma electrons, the beam particles
are treated as macroparticles in PIC algorithm.

\subsection{Low-noise efficient quasi-neutral particle-in-cell code}

The logic behind the quasi-neutral code is that the full dynamics of the
ambient plasma need not be simulated, and its density can be obtained from
the quasi-neutrality condition: 
\begin{equation}
n_{pe}(\vec{x})=n_{0}-n_{b}(\vec{x}).  \label{eq:neutrality}
\end{equation}%
Therefore, ambient plasma is modelled as a passive fluid that responds to
the evolving electron beam in order to maintain charge neutrality. Electron
beam particles are modelled using numerical macro-particles that are
advanced in time by the self-consistently determined electric and magnetic
fields. The leading magnetic field $\vec{B}_{\perp }=-\vec{e}_{z}\times \vec{%
\nabla}_{\perp }\psi $ develops in the $x-y$ plane, where $\psi $ is the $z$%
-component of the vector potential. The inductive electric field associated
with the time-varying flux $\psi $ is $E_{z}=-(1/c)\partial _{t}\psi $.
Electric field also has a transverse component $\vec{E}_{\perp }$ that is
found from the quasi-static force balance of the ambient plasma electrons in
the plane: $\vec{E}_{\perp }+\vec{v}_{pz}\times \vec{B}_{\perp }/c=0$, where 
$\vec{v}_{pz}\equiv v_{pz}\vec{e}_{z}$ is the return flow of the ambient
plasma. This quasi-equilibrium is the consequence of another observation
from direct PIC simulations: that the transverse velocity $\vec{v}_{p\perp }$
of the ambient plasma electrons is considerably smaller than the beam's
average transverse speed $|\vec{v}_{b\perp }|$ and plasma's longitudinal
velocity $v_{pz}$. One of the consequences of that is that the dominant
magnetic field is the transverse one~\cite{EdLee_PoF80}, i.~e. that the
out-of-plane magnetic field is small: $|\nabla \psi |>>B_{z}$. To summarize,
these are the dominant electric and magnetic fields of the quasi-neutral
beam-plasma system: 
\begin{equation}
\vec{B}_{\perp }=-\vec{e}_{z}\times \vec{\nabla}_{\perp }\psi ,E_{z}=-\frac{1%
}{c}\frac{\partial \psi }{\partial t},\vec{E}_{\perp }=-(v_{pz}/c)\vec{\nabla%
}_{\perp }\psi .  \label{eq:major_fields}
\end{equation}

For collisionless plasma, two important conservation laws simplify the
description of the plasma motion: conservation of the canonical momentum in
the $z$-direction and the conservation of the generalized vorticity~\cite%
{buneman_52,kaganovich_pop01,Us}. The former is essential for deriving the
field equation for $\psi $ that defines the dominant in-plane magnetic
field. Conservation of the canonical momentum translates into the
non-relativistic expression for the plasma return velocity: $v_{pz}/c=\tilde{%
\psi}$ (or $v_{pz}/c=\tilde{\psi}/\sqrt{1+\tilde{\psi}^{2}}$ in the
relativistic case), where $\tilde{\psi}=e\psi /mc^{2}$ is the dimensionless
vector potential. From Ampere's law then follows that 
\begin{equation}
\nabla _{\perp }^{2}\psi -4\pi env_{pz}/c=-4\pi J_{bz}/c,
\label{eq:psi_general}
\end{equation}%
where the displacement current is neglected to be consistent with the
quasi-neutrality assumption. The $z$-component of the beam electron current
is calculated from the beam's macro-particles' contribution: $%
J_{bz}=-\sum_{j}q_{j}v_{jz}\delta ^{2}(\vec{x}-\vec{x}_{j})/L_{z}$, where
index $j$ labels numerical macroparticles, and $q_{j}/L_{z}$ and $%
M_{j}/L_{z} $ are the macro-particle's charge and mass per unit length. For
non-relativistic collisionless plasma electrons Eq.~(\ref{eq:psi_general})
is simplified to $\nabla _{\perp }^{2}\psi -k_{pe}^{2}\psi =-4\pi J_{bz}/c$,
where the spatially-nonuniform $k_{pe}^{2}\equiv \omega _{pe}^{2}/c^{2}$ is
obtained from Eq.~(\ref{eq:neutrality}) through $\omega _{pe}^{2}(\vec{x}%
)=4\pi e^{2}n_{e}(\vec{x})/m$. Specifically, $\vec{\nabla}_{\perp }\times 
\vec{J}_{b\perp }$ is small in the linear limit $n_{b}<<n_{p}$

As in the standard PIC, beam electrons are modelled kinetically using
macro-particles with the effective per-unit-length charges and masses $q_{j}$
and $m_{j}$ satisfying $q_{j}/m_{j}=e/m$, where index $j$ labels numerical
macro-particles. The longitudinal momentum of a beam electron (assumed
collisionless owing to its relativistic energy) is found from the
conservation of the canonical momentum: 
\begin{equation}
\gamma _{j}v_{jz}=\gamma _{j0}v_{jz0}+\frac{e}{mc}(\psi -\psi _{j0}),
\label{eq:momentum_cons}
\end{equation}%
where we assume that the initial field in the plasma vanishes: $\psi _{j0}=0$%
. The transverse equation of motion for the beam electrons is: 
\begin{equation}
\frac{d(\gamma _{j}\vec{v}_{j\perp })}{dt}=-\frac{e(v_{jz}-v_{pz})}{mc}\vec{%
\nabla}_{\perp }\psi ,  \label{bm_tr1}
\end{equation}

where the second term in the rhs of Eq.~(\ref{bm_tr1}) is due to the extra
pinching of the electron beam provided by the transverse ambipolar electric
field $\vec{E}_{\perp }=-(v_{pz}/c)\vec{\nabla}_{\perp }\psi $ that develops
in response to the $\vec{v}\times \vec{B}$ expulsion of the electron plasma
fluid. Note that $\vec{E}_{\perp }$ \textit{counters} the magnetic expulsion
of the ambient plasma, yet \textit{reinforces} the magnetic pinching of the
beam. The above expression for $\vec{E}_{\perp }$ are only valid in the
absence of the complete plasma expulsion from the beam filaments, and needs
to be modified when such expulsion takes place. 

\section{Stages of the Weibel Instability}

\begin{figure}[t]
\includegraphics[height=0.15\textheight,width=.5\columnwidth]{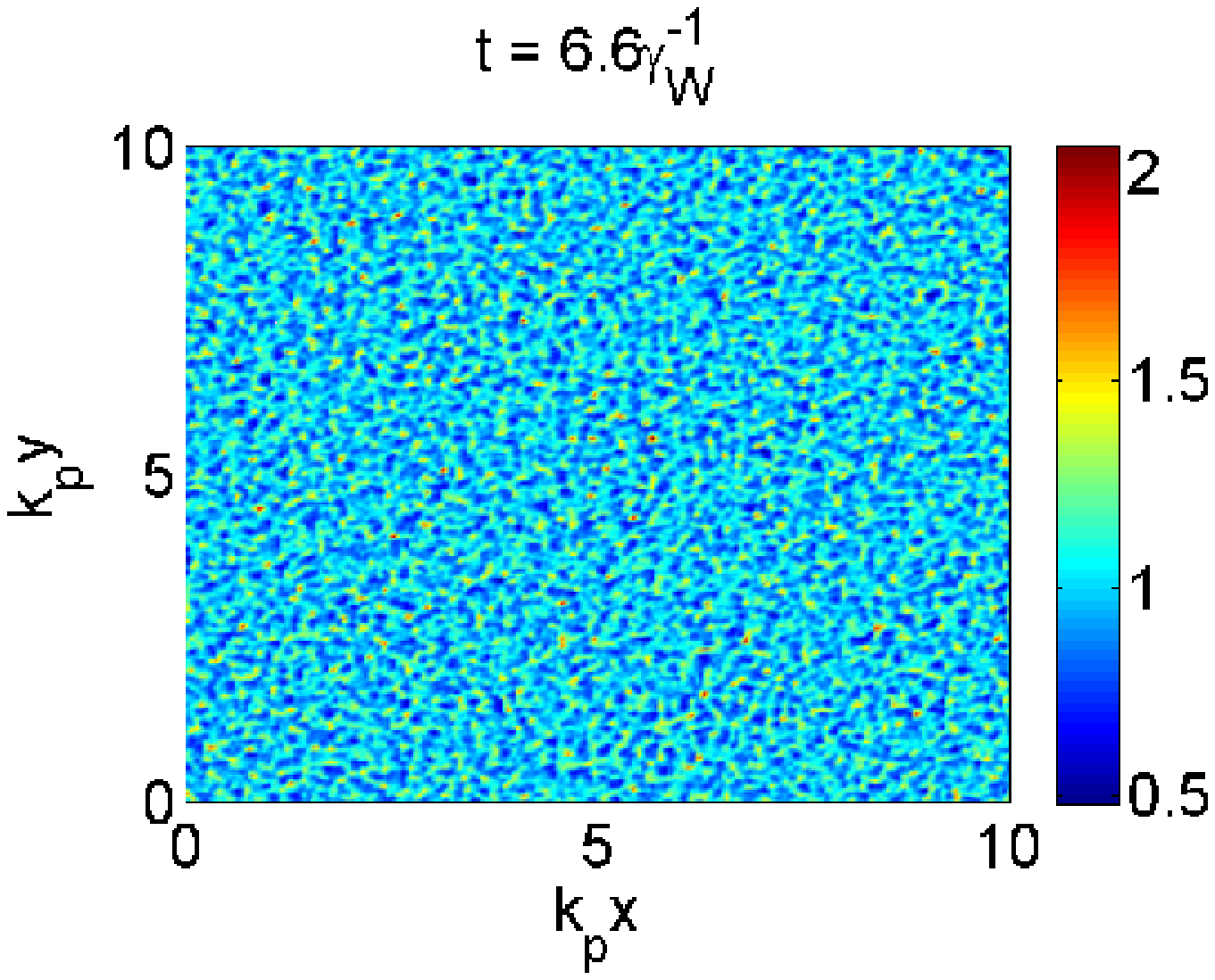}%
\includegraphics[height=0.15\textheight,width=.5\columnwidth]{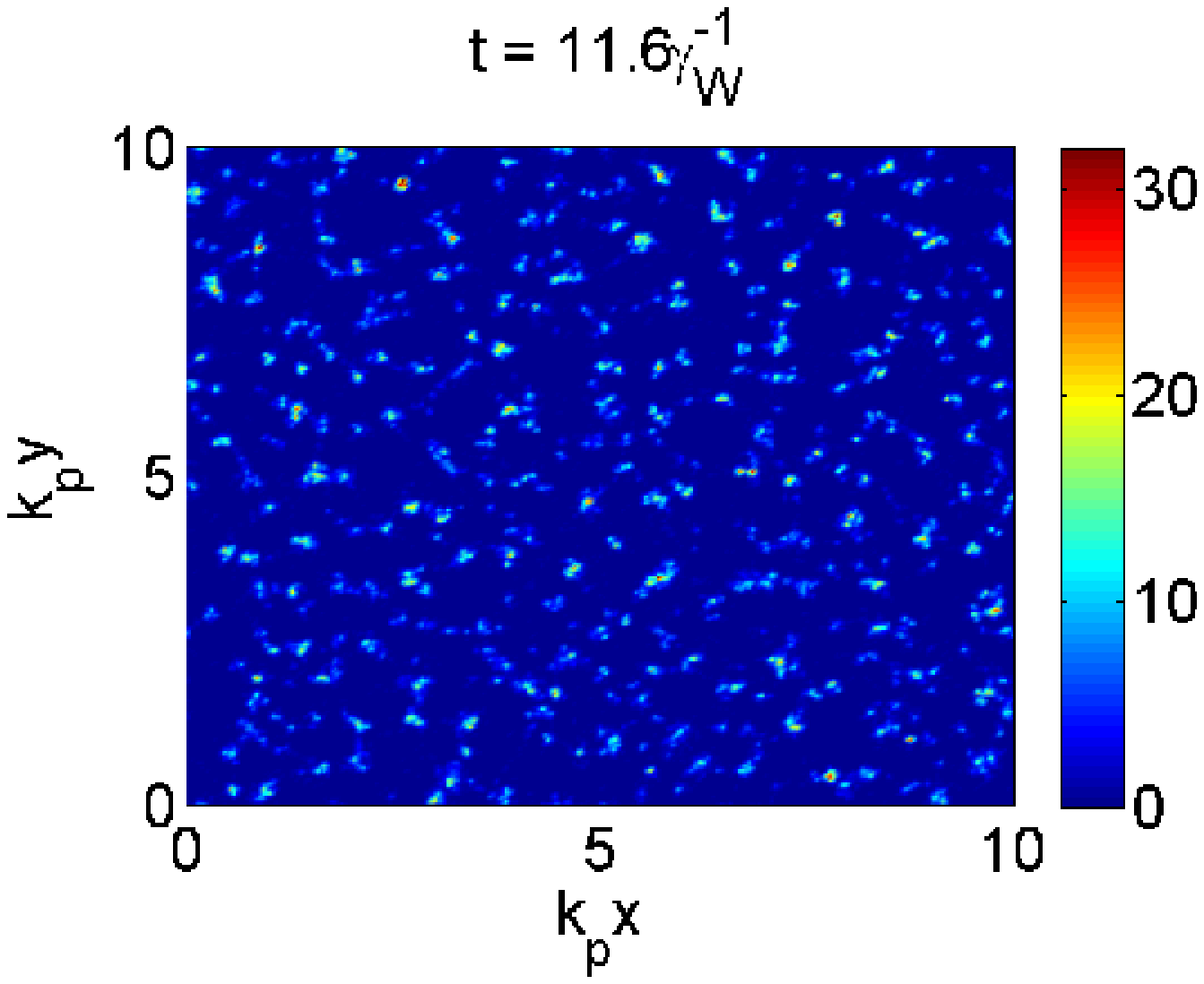} %
\includegraphics[height=0.15\textheight,width=.5\columnwidth]{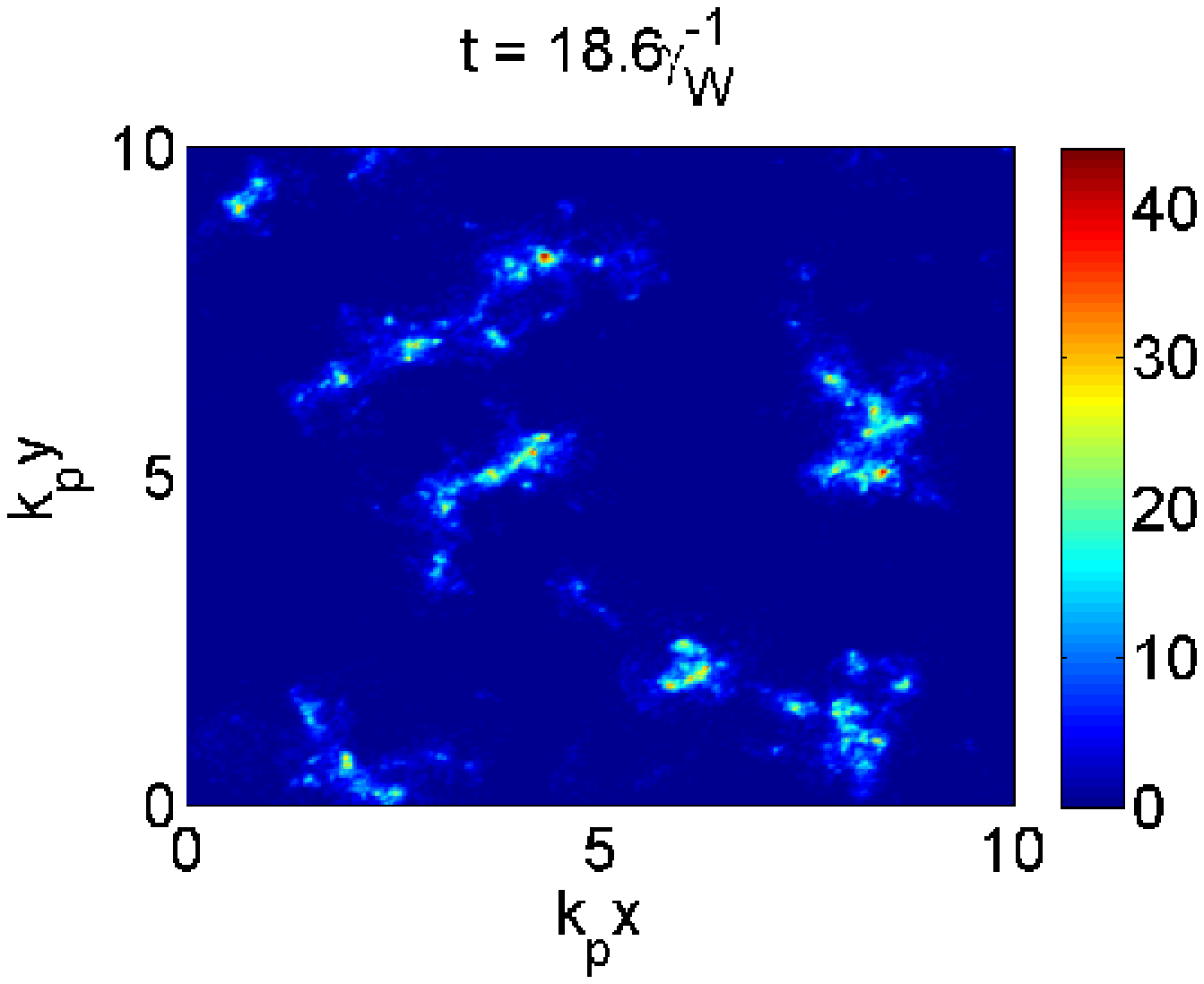}%
\includegraphics[height=0.15\textheight,width=.5\columnwidth]{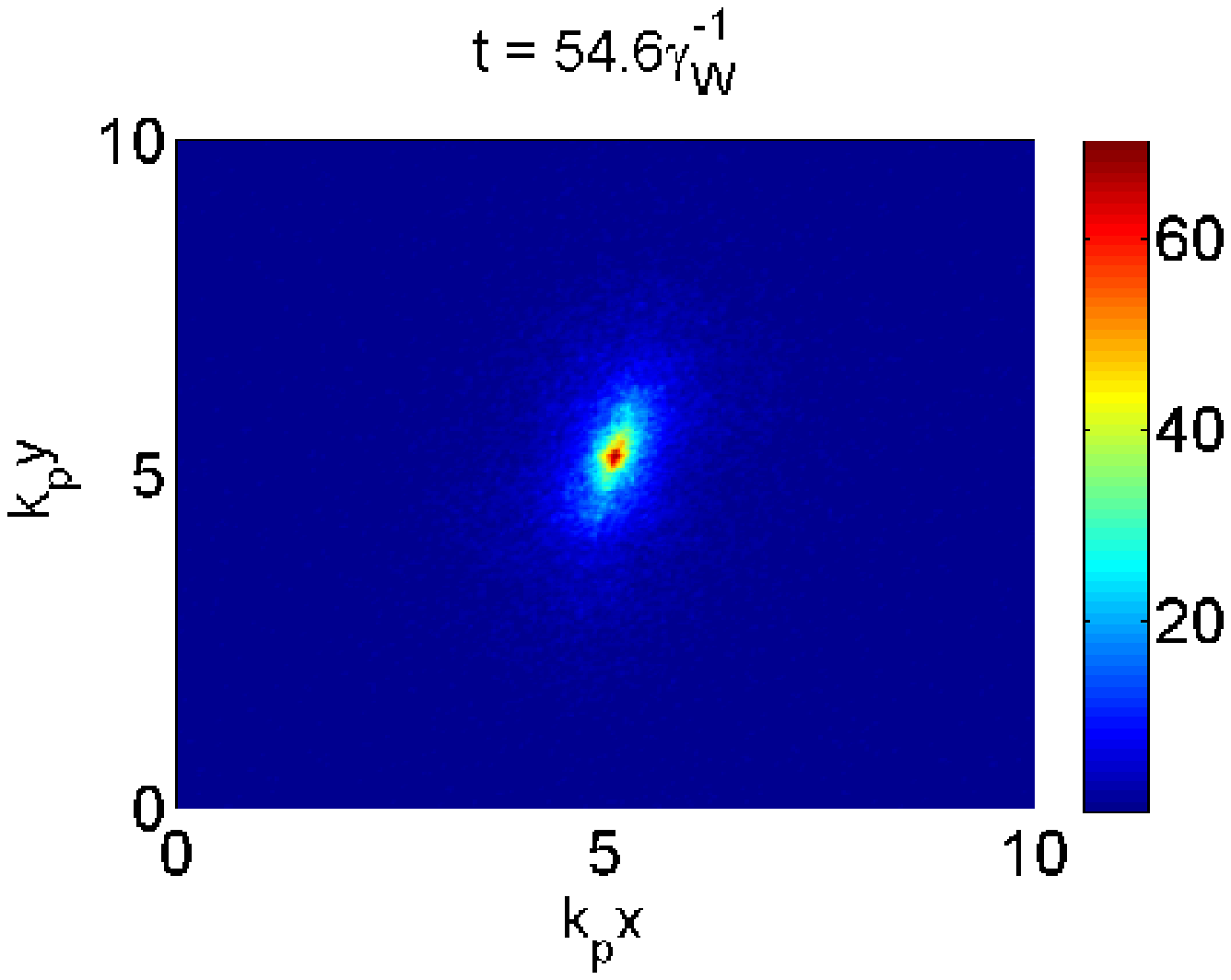}
\caption{The beam density $n_b/n_{b0}$ in the beam cross-section at different
moments of time. First, the beam breaks into a large number of non-stationary filaments in which fluctuations of the beam density are of the order of $n_{b0}$. Non-stationary filaments eventually coalesce into one quasistationary filament.}
\label{1Dpot1}
\end{figure}

During the extensively studied~\cite{Morse,David,Lampe,Medv,Mor1,Mor2,Us}
 early stage of Weibel instability, the  electron beam of the density 
$n_{b0}<n_{p}$ and the radius $R_{\mathrm{init}}\gg c/\omega _{pe}$ breaks up into a large number of filaments, see Fig.~\ref{1Dpot1}.  We model low-density electron beams  with
large cross-section radius  by
performing simulations in square-shape domain  using periodic boundary
conditions (in the simulations, we chose $n_{b0}/n_p=0.001$ and  $k_pL=10$). Small random perturbations are imposed  initially on the
homogeneous beam density. The beam temperature is assumed to be small, the relativistic factor $\gamma_{b0}=10$,  and the beam current is  compensated by the plasma return current. 
\begin{figure}[t]
\includegraphics[height=0.11\textheight,width=.5\columnwidth]{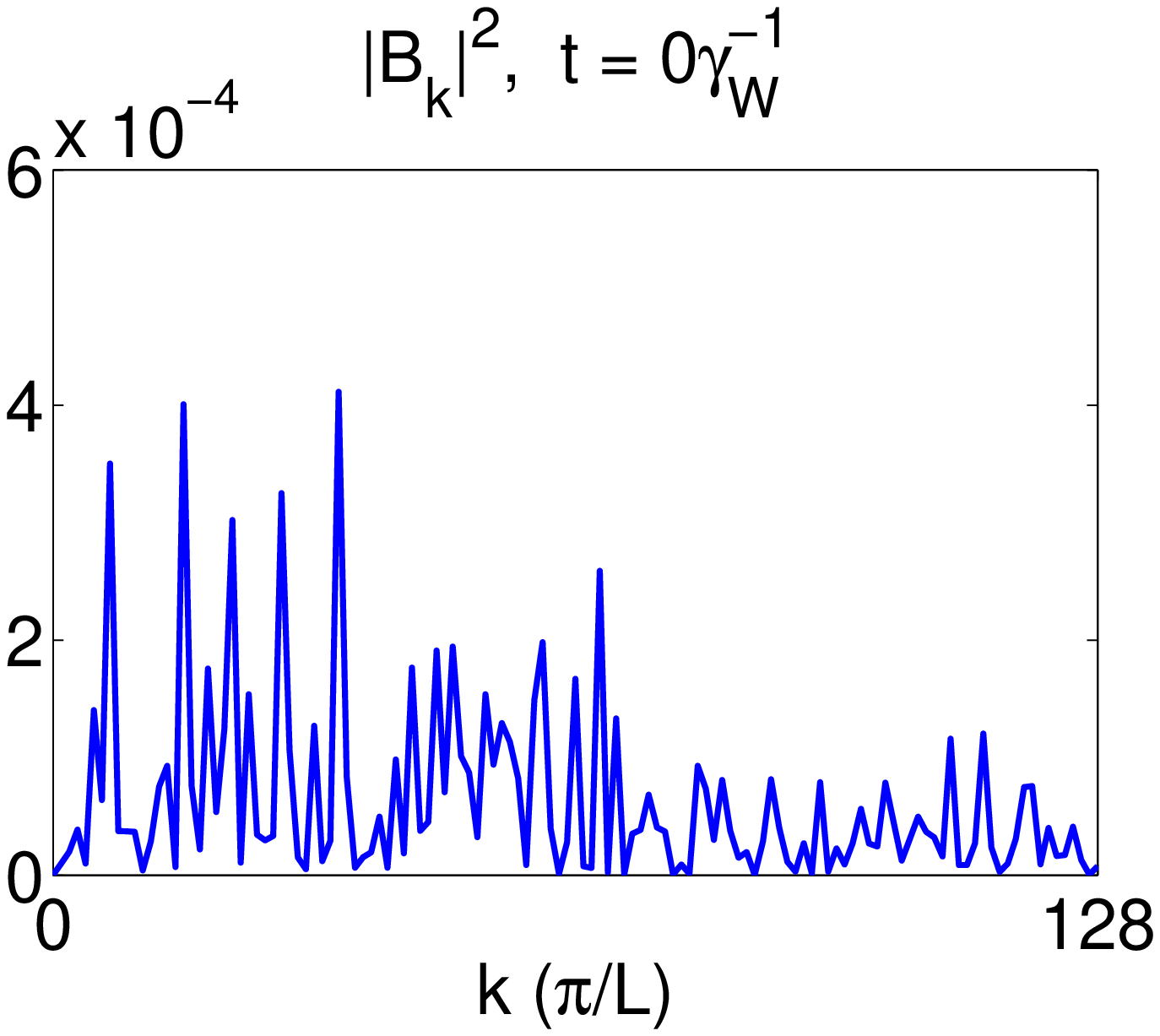}%
\includegraphics[height=0.11\textheight,width=.5\columnwidth]{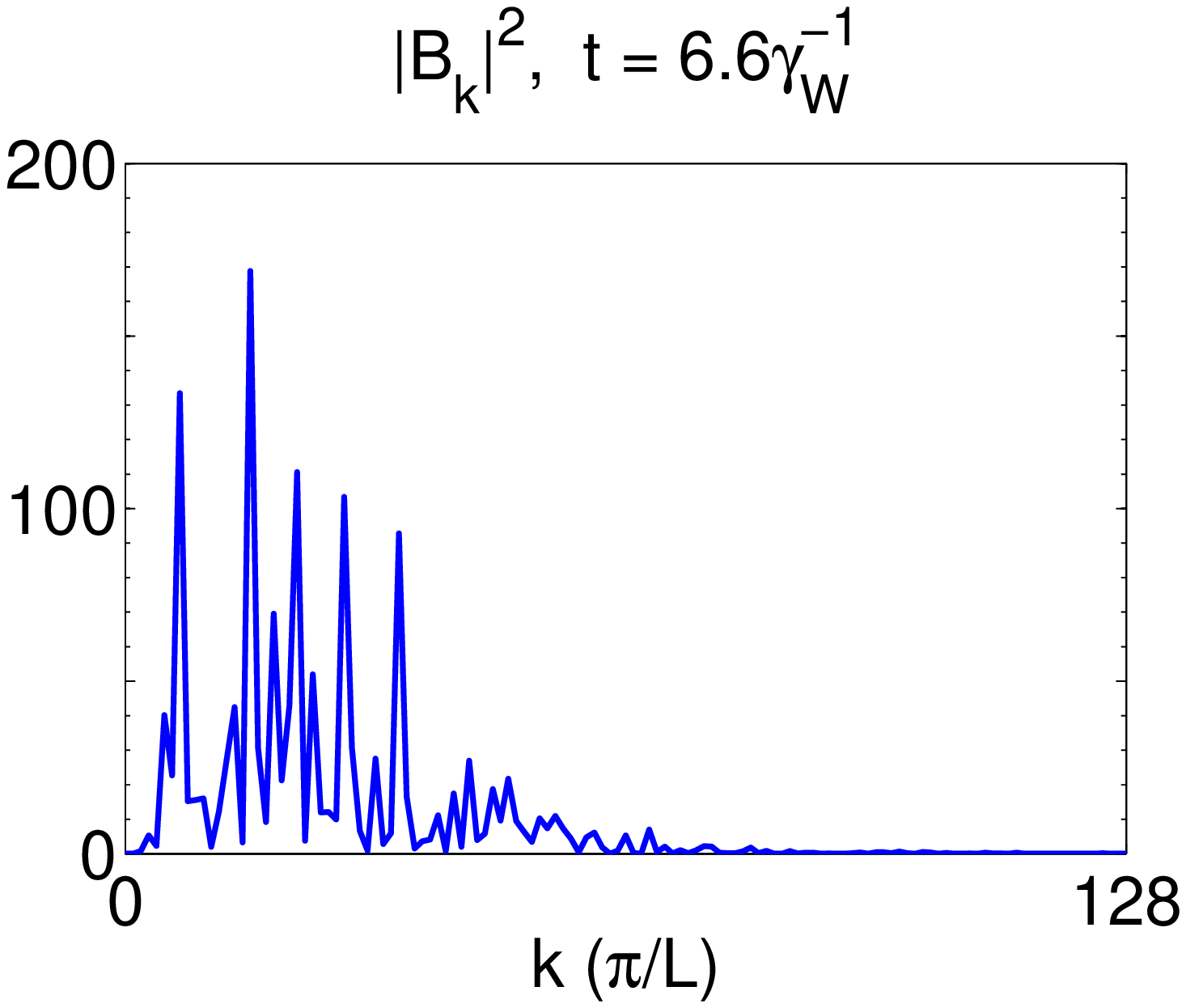} %
\includegraphics[height=0.11\textheight,width=.5\columnwidth]{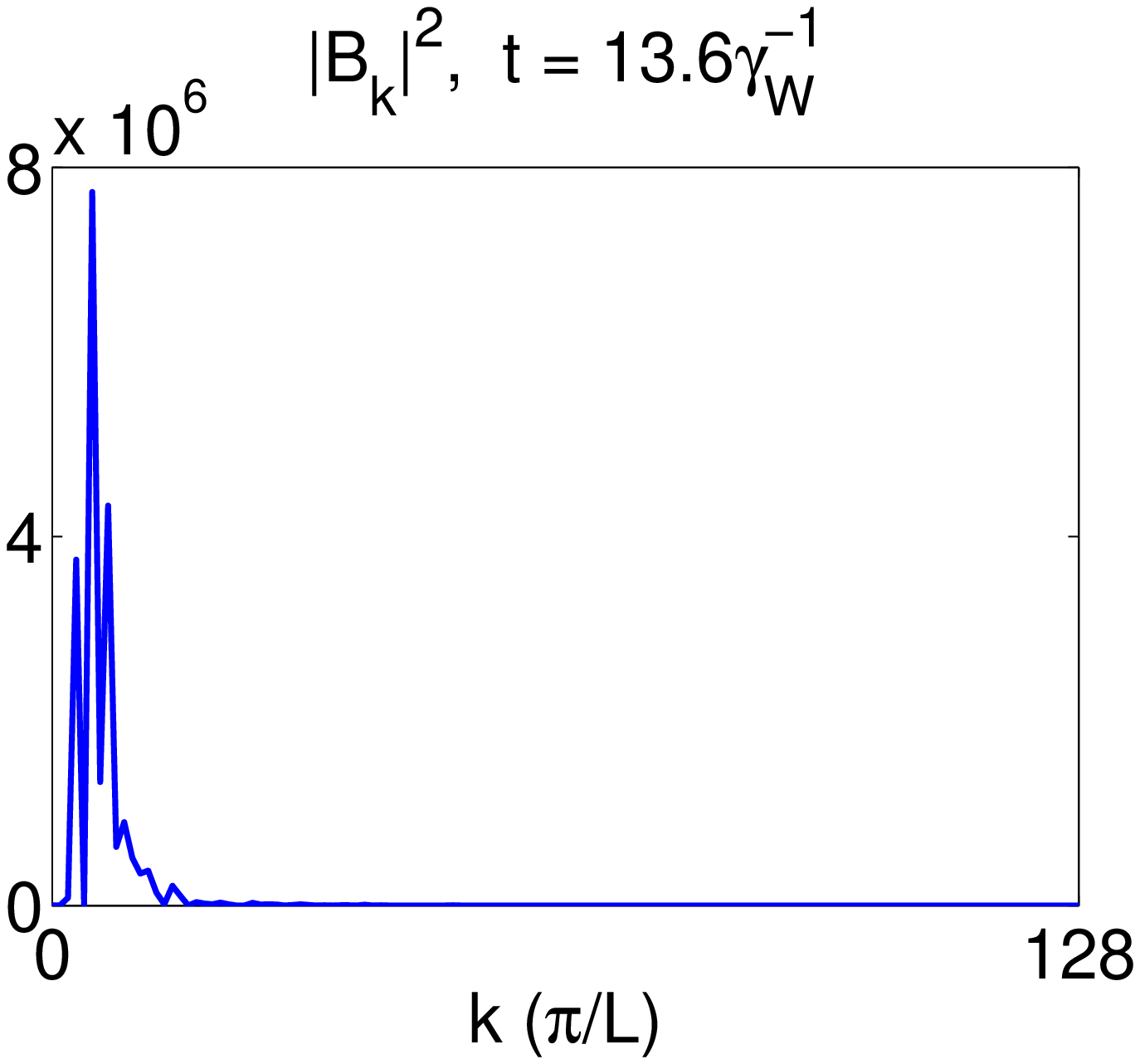}%
\includegraphics[height=0.11\textheight,width=.5\columnwidth]{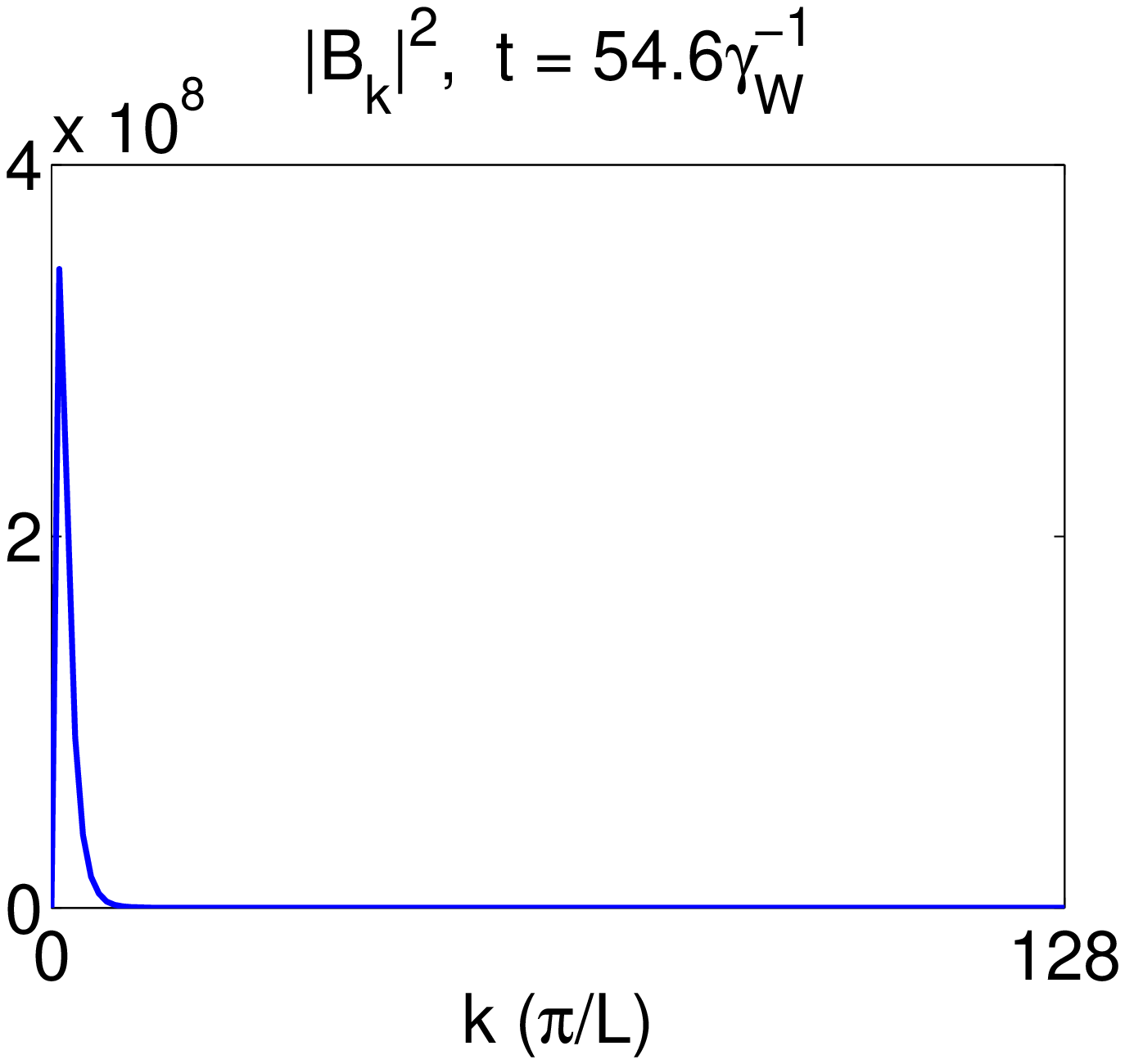}
\caption{Evolution of the magnetic energy spectrum of perturbations. Growth of short-wave perturbations is suppressed due while long-wave perturbation  grow until the system reaches equilibrium.}  
\label{fig:spectrum}
\end{figure}
\begin{figure}[t!]
\includegraphics[height=0.15\textheight,width=.75\columnwidth]{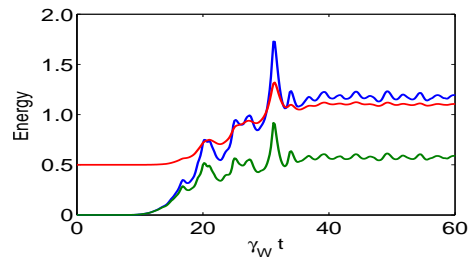}
\caption{The evolution of  the transverse kinetic energy of beam particles(blue),   the magnetic energy (green), and  the energy of plasma particles (red). The saturation level are of  Non-zero initial value of the  plasma energy  is determined by the  energy in the return current. The energy is measured in units $N T_*$, where $N$ is the number  particles in the beam and $T_*=mc^2 n_{b0}/n_p$.}
\label{1dEnergy}
\end{figure}
\begin{figure}[t]
\includegraphics[height=0.18\textheight,width=.75\columnwidth]{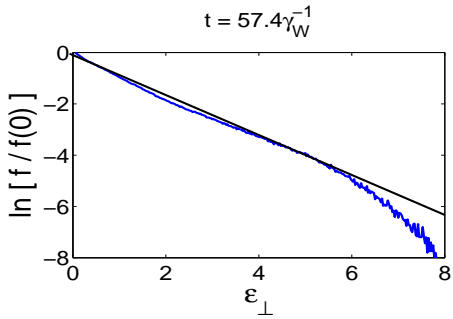}
\caption{The beam particle distribution $\ln [f(\protect\varepsilon _{\bot
})/f(0)]$ over  transverse kinetic energy $\protect\varepsilon_{\bot }$ (measured in units $T_*$). The transverse temperature extracted from this dependence $T\approx 1.2 T_*$.}
\label{Maxwellian}
\end{figure}
 When perturbations are small, the magnetic field energy grows exponentially with the
growth rate of order $\gamma _{\mathrm{WI}}$ given by 
\begin{equation}
\gamma _{\mathrm{WI}}(k_{\perp })=\omega _{p}\sqrt{\frac{n_{b}}{\gamma _{b}n_{p}}\frac{%
\beta _{z}^{2}k_{\perp }^{2}}{k_{\perp }^{2}+k_{p}^{2}}}.
\end{equation}
where  $\beta _{z}=v_{bz}/c$ and $\vec{k}_{\perp }$ is the transverse component  of the wavevector. Although all modes with $k_{\perp }>>k_{p}$ grow simultaneously (with the rate $\gamma _{\mathrm{W}}=\gamma _{\mathrm{WI}}(k_{\perp }\rightarrow\infty)$), wavebreaking occurs first in short-wave modes and then in long-wave modes. Subsequent formation of multistream flow in the transverse beam motion  effectively increases the transverse temperature of the beam particles. The thermal motion, in turn, suppresses growth of small-scale perturbations while large scale perturbations continue to grow. The dynamics of this complex process is  illustrated in Fig.~\ref{fig:spectrum} where the  evolution of the magnetic energy spectrum of perturbations is presented. The evolution of the transverse kinetic energy of beam particles and the magnetic energy is given in Fig.~\ref{1dEnergy}. In the course of multiple merging, non-stationary filaments become bigger and the distance between them increases. Eventually, they coalesce in one quasistationary filament (in our example, it happens at $t\sim 31\gamma_W^{-1}$). Often, this filament has elliptical form and  slow rotation is observed. Particles in the filament perform complicated oscillatory motion along nonclosed trajectories. Nevertheless, our simulations show that the  distribution function of the trapped particles  over the transverse kinetic energy is approximately Maxwellian, see Fig.~\ref{Maxwellian}. Apparently, Maxwellization is caused by multi-generation merging of filaments.  

Note  that due to wavebreaking, fluctuations of the beam density  become  significant (of the order of $n_{b0}$) almost from the very beginning.  
As seen from Fig.~\ref{1Dpot1},  the density in the quasistationary filament exceeds tens times the initial level of the density. Although it is not noticeable in the figure, there is
a big fraction of beam particles (in our case about $30\%$) outside of the
filament in the large area where the final beam density is small, $n_{b}<n_{b0}$.
These particles get dispersed in the process of multiple filament merging, and gain sufficient transverse temperature    forming stable (against further development of Weibel instability)  background . 

\section{Estimates  of the quasistationary filament parameters. }

Characteristics of the quasistationary filament can be estimated by the
following way. Making a rough assumption that all particles from $L\times L$
area are gathered in small area of the radius of the order of skin-depth
length $\delta_p=k_{p}^{-1}$, one can estimate the beam density there as $n_{b}\sim
n_{b0}\ast k_{p}^{2}L^{2}$ and the magnetic field as $B\sim 4\pi en_{b}/k_{p}
$, so that $B^2\sim 4\pi mc^2 n_b^2/n_p\sim 4\pi mc^2 (n_{b0}^2/n_p)k_p^2L^2$. Equating the transverse thermal pressure in the filament to the pressure
of the magnetic field, $n_{b}T_{\perp }\sim B^{2}/8\pi $, one can estimate
the transverse temperature in the beam filament 
\begin{equation}
T_{\perp }\sim
mc^{2}\frac{n_{b0}}{2n_{p}}k_{p}^{2}L^{2}\sim 2 \pi  mc^{2}\frac{I_{b}}{I_{A0}},
\label{eq:T_initial}
\end{equation}%
where $I_{A0}\equiv\beta mc^{3}/e$. This is a very rough estimate. More
accurate formula (\ref{eq:T_final}) which takes into account the spatial
distribution of beam particles in the filament gives considerably smaller
numerical coefficient. The fraction of the  longitudinal kinetic beam energy, $%
W_{||}=\gamma _{b}mc^{2}(n_{b0}L^{2})$, transferred to the transverse beam
motion is given by 
\begin{equation}
\frac{n_{b}(\pi k_{p}^{-2})T_{\perp }}{W_{||}}\sim (k_{p}L)^{2}\left( \frac{%
\gamma _{W}}{\omega _{p}}\right) ^{2},
\end{equation}%
The energy transferred to the magnetic field and to the plasma is of the
same order. Since the beam density $n_{b}$ in our consideration is always
less than $n_{p}$, filaments of the radius $c/\omega _{p}$ carry
sub-Alfvenic currents, $en_{b}c\beta _{z}\pi c^{2}/\omega _{p}^{2}<<\beta
_{z}mc^{3}/e$.

\section{Similarity and conservation laws for the beam filamentation dynamics in the limit of
low density}

When $n_{b}<<n_{p}$ during entire beam evolution and $\vec{V}_{b}\approx
(0,0,-c)$, for analytical tractability we introduce further simplification
in our hybrid model. To the first order in the small parameter $n_{b}/n_{p}$, the beam dynamics is governed by the equations:%
\begin{eqnarray}
k_{p}^{-2}\Delta \psi =\psi -(n_{b}/n_{p})mc^{2}/e,  \label{eq:field}\\
m\gamma _{b}\frac{d\vec{v}_{j\perp }}{dt}=e\nabla _{\perp }\psi ,
\label{eq:motion}
\end{eqnarray}
From these equations one can find that the evolution of the low density beam beam obeys a similarity law: it remains the same when
space coordinates, time, the beam density and vector potential rescale as 
\begin{eqnarray}
&\vec{r}_{\perp } \propto \lambda _{p},  \label{eq:scale1} \\
&t^{-1} \propto \omega _{p}\sqrt{n_{b0}/\gamma _{b}n_{p}}\sim \gamma _{W},
\label{eq:scale2} \\
&\vec{v}_{\perp }\propto c\sqrt{n_{b0}/\gamma _{b}n_{p}}\sim c(\gamma
_{W}/\omega _{p}),  \label{eq:scale3} \\
& n_{b} \propto n_{b0},  \label{eq:scale4} \\
&\psi  \propto (n_{b0}/n_{p})mc^{2}/e,  \label{eq:scale5}\\
&\frac{\gamma_b mv_{\perp}^2}{2}\propto mc^2 n_b/n_p.\label{eq:scale6}
\end{eqnarray}
\textbf{In particular, it means that at small beam densities it is enough to
study beam propagation only at the one level of the initial density}. The
propagation at other levels can be found by rescaling the simulation results
according to Eqs.~(\ref{eq:scale1})-(\ref{eq:scale5}). As seen from Eqs.~(%
\ref{eq:scale2}), (\ref{eq:scale3}), (\ref{eq:scale5}) and (\ref{eq:scale6}), big beam
relativistic factors $\gamma _{b}$ slow down the beam filamentation not
changing magnetic fields created by filaments and transverse kinetic energy of beam particles.

It follows from the conservation of longitudinal momentum that for ultrarelativistic beam 
$\gamma_b mc^2\approx \gamma_{b0}mc^2+m \gamma_{b0}m v_{\perp}^2/2-e(\psi-\psi_*)$, where $\psi_*$ is the initial value of the magnetic potential. The energy conservation law ~\cite{Karmakar_}  can be now transform to the form 
\begin{equation}  \label{eq:energy_con}
W_{\perp}=\frac{1}{2}\int n_b m\gamma_b v_{\perp}^2dS-\frac{1}{2}\int e n_b\psi
dS=const.
\end{equation}
This formula can be also derived directly from Eqs.~(\ref{eq:field}) and (\ref{eq:motion}). The potential energy is given by the second term in the right-hand side of Eq.~(\ref{eq:energy_con}); it includes the energy of the magnetic field and the plasma
motion:
\begin{equation}  \label{eq:pot_energy_con}
\frac{1}{2}\int e n_b\psi
dS=\frac{1}{2}\int B^2dS+\frac{1}{8\pi}\int n_pmv^2dS,
\end{equation}
where $\vec{B}=\nabla \psi$ and $v=e\psi/mc$. 
  The coefficient $1/2$ in front of the potential term 
reflects the fact that each particle is counted twice in the integral (\ref%
{eq:energy_con}) so that the total energy of the one particle (which in general
varies with time!) is 
\begin{equation}
\epsilon_{tot}=\frac{\gamma _{b}m v_{\perp }^{2}}{2}-e\psi.  \label{eq:energy_part}
\end{equation}
The magnetic potential  forms a potential wells for particles, $U=-e\psi$; these wells merge with each other when corresponding filaments merge. 

The other obvious integral on the motion of the system (\ref{eq:field})-(\ref%
{eq:motion}) is the conservation of the total number of the macroparticles: 
\begin{equation}  \label{eq:particle_con}
N=\int n_b dS=const.
\end{equation}
Note that for the beam with initially negligible transverse velocity spread
homogeneously distributed over the area $L\times L$, the magnetic potential $%
\psi=(n_b/n_p)mc^2/e$ and the total energy and number of particles are given
by formulas, 
\begin{equation}  \label{eq:inital_valuesNW}
W_{\perp}=-\frac{1}{2}Ne\psi_*=-\frac{1}{2}NT_*, \quad N=n_{b0}L^2,
\end{equation}
where $e\psi_*=T_*=(n_{b0}/{n_p})mc^2$.

It is instructive to note that, in the framework of Eqs.~ (\ref{eq:field}) and (\ref{eq:motion}),  the magnetic potential can be presented as a sum of potentials created by each beam particle,     
\begin{equation}
\psi(\vec{r},t) =\frac{mc^2}{2\pi e n_p}\int K_0(k_p|\vec{r}-\vec{r'}|)n_b(\vec{r}',t)d\vec{r}'.
\label{eq:Bessel}
\end{equation}
Thus, the interaction between particles in the  low density beam  is a pairwise Coulomb attraction  screened by Bessel function at distances larger than the skin-depth $k_p^{-1}$, and the potential energy of beam particles can be presented as  a sum of potential energies of all particle pairs.

\section{Structure of filaments }

\begin{figure}[t]
\includegraphics[height=0.11\textheight,width=.51\columnwidth]{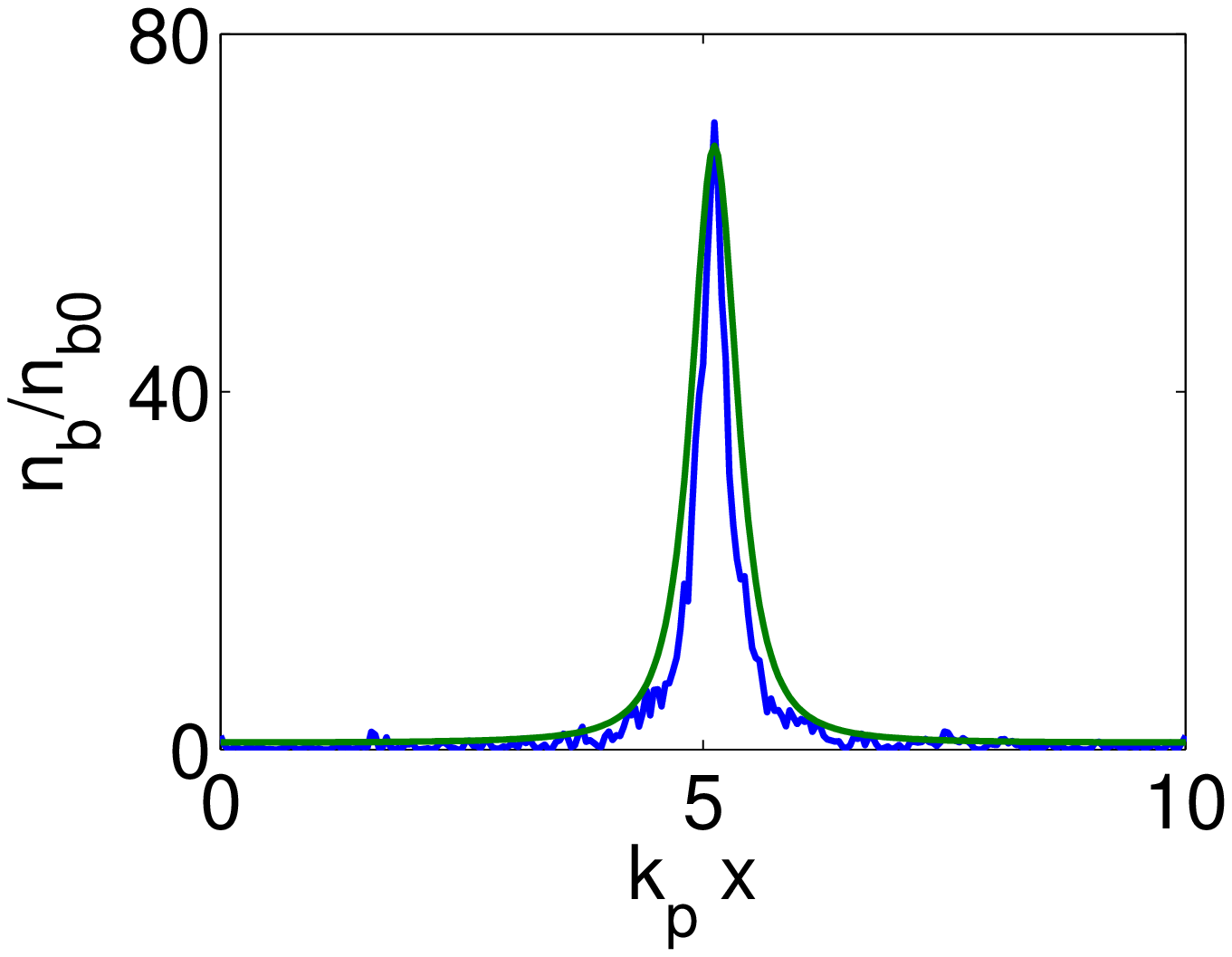}%
\includegraphics[height=0.11\textheight,width=.51\columnwidth]{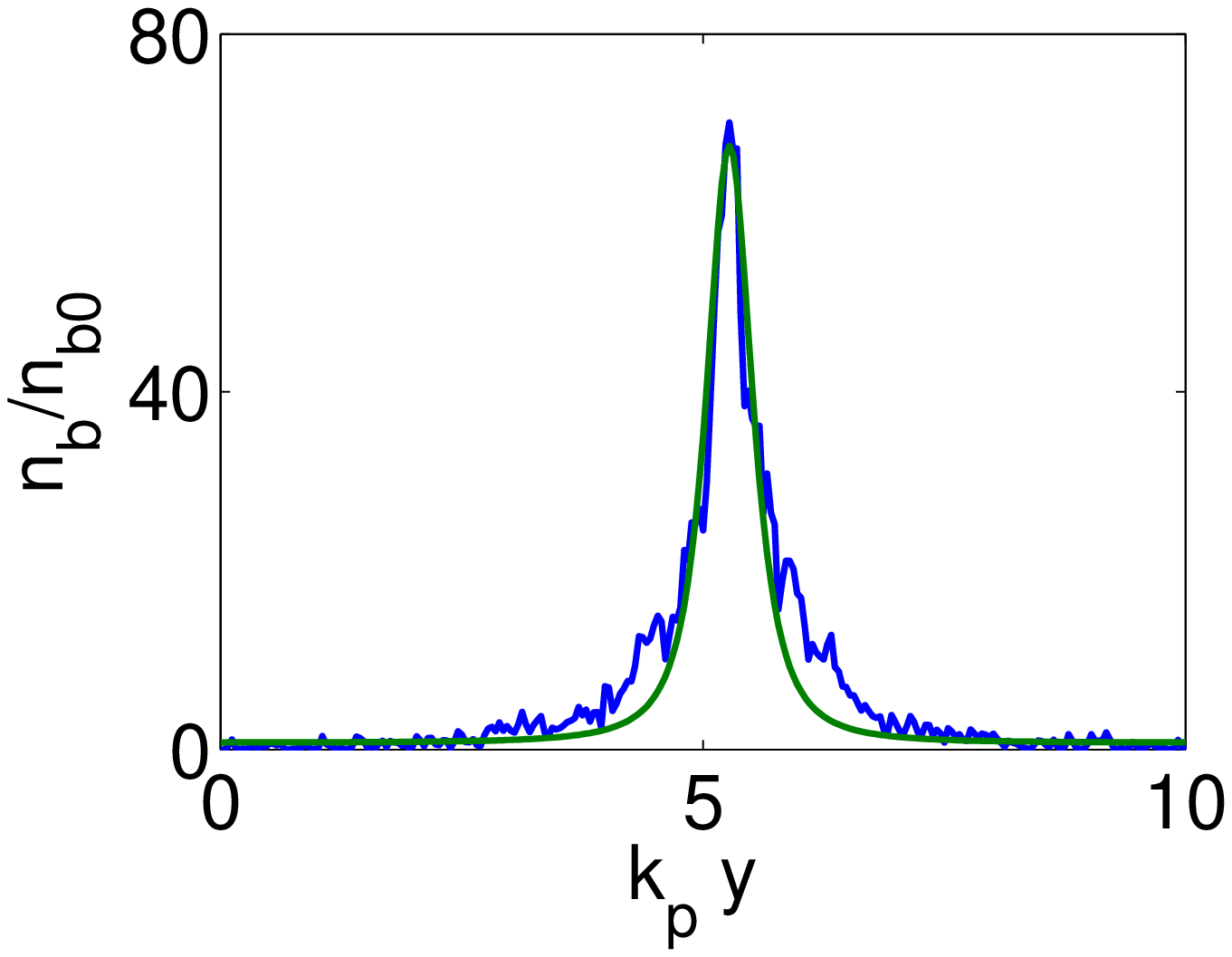} %
\caption{ Crosssections of the filament density (blue) in $x$ and $y$ directions at the moment $t\approx 54 \gamma_W^{-1}$. Green lines correspond to Bennett pinch  with the radius $r_B=0.4k_p^{-1}$. The temperature of the filament is taken from simulations, $T=1.35 mc^2n_{b0}/np$.}  
\label{fig:comparison_density}
\end{figure}

Figure \ref{fig:comparison_density} shows vertical and horizontal density cross-sections of the beam corresponding to the density distribution in the quasistationary filament in Fig.~\ref{1Dpot1}.  As can be seen from this figure the filament pinched to the radius even smaller than the skin depth $\delta _{p}=k_p^{-1}$. For such radius the plasma screening of the magnetic
field does not occur. Therefore, the magnetic pinching force is balanced by
the pressure gradient $\nabla p=j\times B$ and $\nabla \times B=4\pi j/c$.
This equilibrium correspond to the Bennett pinch \cite{DavBook}
\begin{equation}
n_{b}=\frac{8/(k_{p}r_{B})^{2}}{(1+(r/r_{B})^{2})^{2}}\frac{T}{mc^{2}}n_{p},
\end{equation}

\begin{figure}[h]
\includegraphics[height=0.15\textheight,width=.71\columnwidth]{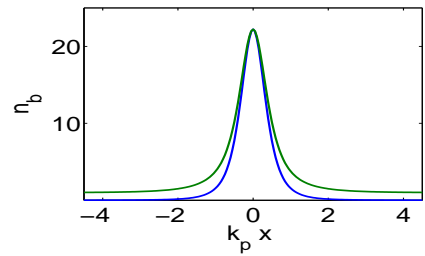}
\caption{The beam particle density in Bennett pinch (red) and modified
Bennett pinch (blue), $k_{p}r_{B}=0.6$.}
\label{BennettPinch}
\end{figure}

For the Bennett pinch the self-magnetic field outside of the filament is
decreasing with radius $B=4\pi I/cr$ and magnetic potential $\psi $
decreases as $-4\pi (I/c)\ln r$. Therefore according to the Boltzmann
relationship the density $n_{b\ast }\exp (e\psi/T)$ decreases as power law of
radius. Due to plasma screening of the magnetic field, the magnetic field
and variation of the magnetic flux vanishes at $r>>\delta _{p}$. Therefore
density does not approach zero at large radius but tends to a finite value.
This means that a low density halo forms outside filament. Detailed analysis
shows that the total number density of particles in the
halo is comparable to the total number density in the filament. See detail
calculation in Appendix. 
Difference between Bennett distribution and modified Bennett distribution is
illustrated in Fig. \ref{BennettPinch}. 

\begin{figure}[t]
\includegraphics[scale=0.4]{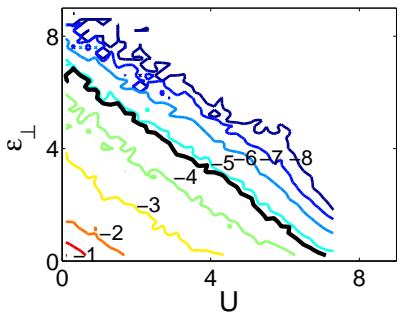}
\caption{Contour lines of the beam particle distribution $\ln [f/f(0)]$ in the  space formed by the transverse kinetic energy $\varepsilon_{\perp}=\gamma_b m v_{\perp}^2/2$ and potential energy $U=-e\psi+e\psi_{max}$ (measured in units $T_*=mc^2n_{b0}/n_p$). The variable $U$ changes from zero (in the center of the filament, where $e\psi =e\psi_{max}\approx 7.6 mc^2n_{b0}/n_p$) to the value $7.3mc^2n_{b0}/n_p$ outside of the filament (where $e\psi =e\psi_{min}\approx 0.25 mc^2n_{b0}/n_p$). Separatrix (black line) divides the   phase space into two regions: the first region (in the shape of triangle) corresponds to particles trapped in the filament and the second one to untrapped particles. The point $\varepsilon_{\perp}=0$, $U=0$ corresponds to the cold particles at the center of the filament.}
\label{EDFtotal}
\end{figure}

The energy distribution of beam particles   
in the phase space space ($\varepsilon_{\perp},U$) obtained from our simulations is shown in Fig.~\ref{EDFtotal}. (For sake of convenience, the potential energy here is defined as $U=-e\psi+e\psi_{max}$.) One can see from this figure that the distribution function depends approximately only on the total particle energy $\varepsilon_{tot}=\varepsilon_{\perp}+U$. Such dependence is a simple consequence of the phase mixing of particles moving  in the quasistationary potential well along trajectories with close total energies. 
\begin{figure}[t]
\includegraphics[height=0.18\textheight,width=.75\columnwidth]{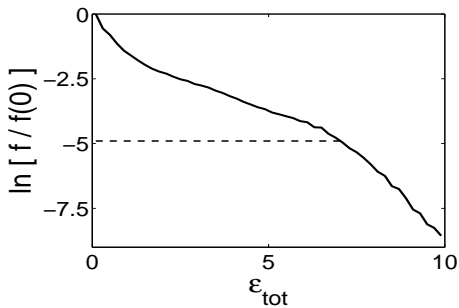}
\caption{The beam particle distribution $\ln [f/f(0)]$ as a function of $\varepsilon_{tot}$ (measured in units $T_*$). Dashed line corresponds to the distribution function when energy exchange between electrons is absent.}
\label{Gurevich}
\end{figure}

The electron distribution as a  function of $\varepsilon_{tot}$ is shown in  Fig.~\ref{Gurevich}.
Note that if there is no energy exchanged between electrons,
the beam particle distribution $f$ over total energy $\varepsilon _{tot}$
trapped in a potential well  is constant ~\cite{Gurevich}. 
Because of the many events leading to the
energy exchange during filament merger the energy electron distribution function is close to a Maxwellian (for ideal Maxwellian distribution function, $\ln [f/f(0)]$ should correspond to the straight line). 
However, there is significant departures from a Maxwellian for low and high
total energy (deeply trapped and untrapped electrons). This because
these electrons do not undergo many energy transfer with other electrons
during filament merger. Low energy electrons are confined to the center of
the filament and do not experience the time dependent magnetic field and
inductive electric field during filament merger. Similarly untrapped
electrons do not participate in energy transfer. Therefore, tail of the EEDF
is strongly depleted at $\varepsilon _{tot}>\psi _{\max }-\psi_{\min}$ . 
%
\section{Energy transfer from the beam longitudinal kinetic energy to the
beam transverse kinetic energy self magnetic field and plasma electrons}
From two conservation laws, conservation of the number of particles and the
energy conservation law, one can find the parameters $T$ and $r_B$ of the modified Bennett pinch and then the energy transfer from the beam longitudinal
kinetic energy to the beam transverse kinetic energy,  magnetic field and
plasma electrons can be calculated. 

Using the Boltzmann distribution of the density in quasistationary state, the conservation of the number of particles gives 
\begin{equation}
n_{b\ast }\int \!\exp (e\psi /T)\,dS=N,\label{eq:particle_con_Max}
\end{equation}%
where $N=n_{b0}L^{2}$. For Maxwellian distribution function,  the energy conservation law (\ref{eq:energy_con}) with the initial constant (\ref{eq:inital_valuesNW}) can be transformed    to 
\begin{equation}
TN=\frac{1}{2}n_{b\ast }\int \exp (e\psi /T)e(\psi-\psi_*) dS,
\label{eq:energy_con_Max}
\end{equation}%
where $\psi_*$ is the initial value of the magnetic potential. The  term in the left-hand side corresponds to the transverse
kinetic energy of the beam particles, and the term in the right-hand side corresponds to the increase in the effective
potential energy.  This increase  is the same as the increase in the energy of the transverse motion (that is,  $<\delta\psi>=T$). Thus, the energy  drawn from  the longitudinal motion of the beam during its evolution is: 
\begin{equation}
<\Delta (\gamma_b mc^2)_{||}>\approx -{2T},\label{eq:beam_energy_loss}
\end{equation}

After substituting the magnetic potential from Eq.~(\ref{eq:interpPSI}) into
Eqs.~(\ref{eq:particle_con_Max}) and (\ref{eq:energy_con_Max}) and performing integrations, one
can find the following asymptotic formula 
\begin{equation}
{T}=  \frac{k_p^2L^2}{1+W(\alpha_0 {k_p^2L^2})}\frac{T_*}{8\pi} =  \frac{ mc^2/2}{1+W(\alpha_0 {k_p^2L^2})}\frac{I_{b}}{I_{A0}},
\label{eq:T_final}
\end{equation}
where $W(x)$ is the product logarithm (or Lambert W-function defined by the equation $W\exp(W)=x$), $\alpha_0=\exp(2\gamma_E-2)/4\pi = 0.0341...$ and 
 $\gamma_E=0.5772...$ is the Euler constant, $T_*=mc^2n_{b0}/n_p$ and $I_{b}/I_{A0}=en_{b0}cL^2/(mc^3/e)$. For $k_pL=10$, Eq.~(\ref{eq:T_final}) gives $T\approx 1.7T_*$while simulations show that the transverse electron energy  averaged over all beam particles is $30\%$ less:  $<\varepsilon_{\perp}>=T\approx 1.2 T_*$. The temperature is higher in the filament $T\approx 1.4 T_*$ and lower outside of it. Difference between theoretical value of $T$ and that obtained from simulations is due to incomplete Maxwellization of the beam electron distribution function.  

Since the product logarithm $W(x)\sim \ln (x/ln x)$ , one can  conclude from Eq.~(\ref{eq:T_final}) that the transverse temperature of the low-density beam slowly decreases  as the beam cros-ssection aria increases (while the beam current is kept constant, $I_b=const$).

At large beam cross-section, the  energy transfered  from the beam longitudinal motion to the beam transverse motion, the magnetic field and the plasma motion is divided  in the following way: 
\begin{eqnarray}
&\hspace{-0.1in} \frac{n_{b0}<\varepsilon_{\perp}>}{2n_{b0}T}=\frac{1}{2}, \label{eq:transverse_energy_gain}\\
&\hspace{-0.1in} \frac{<B^{2}/8\pi>}{2n_{b0}T}\approx \frac{-  \ln r_B+ 1.3 r_B -0.88  }{1+W(\alpha_0 {k_p^2L^2})},
\label{eq:magnetic_energy_gain}\\
&\hspace{-0.11in} \frac{<n_p\Delta \varepsilon_p>}{2n_{b0}T}\approx \frac{1}{2}-\frac{-  \ln r_B+ 1.3 r_B -0.88  }{1+W(\alpha_0 {k_p^2L^2})}.
\label{eq:plasma_energy_gain}
\end{eqnarray}
For  $k_pL=10$, Eq.~(\ref{eq:rB_on_L}) gives $r_B\approx 0.4$  and Eq.~(\ref{eq:magnetic_energy_gain}) gives  ${<B^{2}/8\pi>}/{2n_{b0}T} \approx 0.26$, and Eq.~(\ref{eq:plasma_energy_gain}) gives  ${<n_p\Delta \varepsilon_p>}/{2n_{b0}T}\approx 0.24$. One can find from Fig.~\ref{1dEnergy}  that  the saturation level of the transverse kinetic energy of the beam  ($1.2 NT_*$) is two times higher than the saturation level  of the magnetic energy ($0.6 NT_*$). It is also two times higher than  the saturation level  of  the plasma energy without initial energy ($0.6 NT_*$). This observation is consistent with theoretical prediction.
\section{Conclusion}
We have considered the filamentation process of an electron beam with  small density and large radius propagating through the dense plasma. We have found that the beam dynamics  obeys  simple similarity laws which simplify the beam description. 
During the formation of filaments and their merging the total transverse energy of the transverse motion of the beam particles is conserved. The effective transverse potential energy is comprised mostly of the energy of the magnetic field and the plasma motion and can be presented as a sum of pairwise interaction energies between beam particles.

Our particle-in-cell simulations have shown that all filaments eventually coalesce into one pinched beam although a significant fraction of the particles remains untrapped and scattered outside of this pinch. The electron beam distribution over the transverse kinetic energy is close to the Maxwellian one for the bulk of electrons. We have developed an analytical model and found the distribution of the particles in the modified Bennett pinch  and in the low-density halo around it. In particular, we have found that the radius of the Bennett pinch modified by the return plasma current is always less or of the order of the plasma skin depth. For a given beam transverse size, we have calculated the energy transfer from the beam longitudinal motion to the transverse  kinetic energy of the beam, the self-magnetic field and the plasma motion. It has turned out   that the final transverse beam temperature is proportional to the beam current and slowly  decreases with the beam radius.  The results obtained from the model agree well with those obtained from particle-in-cell simulations.

Note that although our consideration is limited to initially cold beams, the calculations can be performed in a straightforward manner  for the arbitrary case.

\section{\protect\bigskip Appendix I. Detailed characteristics of modified
Bennett pinch}

In this section we describe the structure of an isolated filament similar to
one shown in Fig.~\ref{1Dpot1}. We assume that all particles have Maxwellian
distribution over velocities, i.e. the system is in a thermodynamic
equilibrium. Hence, the density of particles is expected to be distributed
in the effective potential $\psi $ according to Boltzmann law: 
\begin{equation}
n_{b}=n_{b\ast }\exp (e\psi /T),  \label{eq:Boltzmann}
\end{equation}%
where $n_{b\ast }$ is some constant and $T$ is the transverse temperature
(equal to the average transverse energy of a beam particle $T=<m\gamma
_{b}v_{\perp }^{2}/2>$). 

Substitution of Eq.~(\ref{eq:Boltzmann}) into Eq.~(\ref{eq:field}) gives
closed equation for the vector potential 
\begin{eqnarray}  \label{eq:Equilibrium}
k_p^{-2}\Delta \psi= \psi- \frac{mc^2}{e n_p}n_{b*}\exp\Big(\frac{e\psi}{T}%
\Big).
\end{eqnarray}
Since the magnetic filed is small far from the filament, the magnetic
potential should be constant at large distances from the filament. Under
this condition, equation (\ref{eq:Equilibrium}) determines the function $%
\psi(x,y) $ at each value of the parameter $n_{b*}$. 
In one dimensional case when all quantities depend on one transverse
coordinate $x$, one can solve Eq.~(\ref{eq:Equilibrium}) analytically and
find the potential and density distribution in an isolated filament, see Appendix II.

In two dimensional case when particles are gathered from large area $L\times
L$ ( much bigger than the aria occupied by the filament), the fully thermalized filament is axially symmetric and therefore Eq.~(\ref{eq:Equilibrium}) can be transform to
\begin{eqnarray}  \label{eq:axialEquilibrium}
\frac{1}{ k_p^{2}r}\frac{d}{d r} r\frac{d}{d r}\Big(\frac{e\psi}{T}%
\Big)= \frac{e\psi}{T}%
- \frac{1}{k_p^2r_{D}^2}\exp\Big(\frac{e\psi}{T}-\frac{e\psi_0}{T}%
\Big).
\end{eqnarray}
where $r_{D}=[T/4\pi e^2n_b(0)]^{1/2}$ is the Debye radius corresponding to the beam density at the filament center, $n_b(0)=n_{b*}\exp(e\psi_0/T)$, and $\psi_0=\psi(0)$.
 This equation can be solved numerically by the shooting method: at given value of the parameter $k_pr_D$, we  find such $e\psi_0/{T}$ at which
   the  magnetic potential $\psi$
is a monotonically decreasing function of $r$ with asymptote $\psi\rightarrow \psi_{\infty}$ and $d\psi/dr\rightarrow 0$ at $r\rightarrow \infty$. 

It turns out that the solution with these properties exists only when  $k_p r_D<1$. Indeed, at the point $r=0$ where the magnetic potential reaches maximum, we have $\psi=\psi_0$, $d\psi/dr=0$, and $d^2\psi/dr^2< 0$. Therefore, the right-hand side of  Eq.~(\ref{eq:axialEquilibrium}) should be  negative at this point, that is, 
\begin{eqnarray}  \label{eq:inequality_at_center}
\frac{e\psi_0}{T}<
\frac{1}{k_p^2r_{D}^2}.
\end{eqnarray}
 On the other hand, the expression in the right-hand side of   Eq.~(\ref{eq:axialEquilibrium}) should be equal to zero at large $r$. Therefore, 
 \begin{eqnarray}  \label{eq:inequality_at_infinity}
 \frac{e\psi_{\infty}}{T}=
\frac{1}{k_p^2r_{D}^2}\exp\Big(\frac{e\psi_{\infty}}{T}-\frac{e\psi_0}{T}\Big)
>\frac{e\psi_0}{T}\exp\Big(\frac{e\psi_{\infty}}{T}-\frac{e\psi_0}{T}\Big).
\end{eqnarray}
 After simple manipulations, we find that $\psi_{\infty}$ and $\psi_{0}$ satisfy to the following condition,
\begin{eqnarray}  \label{eq:inequality}
F\Big(-\frac{e\psi_{\infty}}{T}\Big)<F\Big(-\frac{e\psi_{0}}{T}\Big)\quad at \quad \frac{e\psi_{\infty}}{T}<\frac{e\psi_{0}}{T},
\end{eqnarray}
where $F(x)\equiv -x\exp(-x)$. Since this function grows only at $x>1$, $F'(x)=(x-1)\exp(-x)$, the condition (\ref{eq:inequality}) imposes restriction on the maximum value of the magnetic potential, $e\psi_0/T>1$. It means that  filament-like solutions exist only when $1/{k_p^2r_{D}^2}>e\psi_0/T>1$, that is, only when $k_pr_D<1$.  Note that the trivial solution of Eq.~(\ref{eq:axialEquilibrium}) $e\psi/T=const=1/k_p^2 r_D^2$ exists at all values of  $k_p r_D$. However, when $k_p r_D\geq 1$, there are no other solutions except the trivial one: the Weibel instability is suppressed by the thermal  motion of particles, and the beam  remains homogeneous with $n_b=const$.

\begin{figure}[h!]
\includegraphics[scale=0.6]{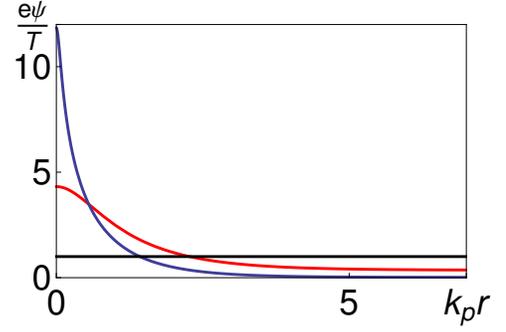}
\caption{The magnetic potential at different values of the Debye radius: $k_pr_D=1$ (black), $k_pr_D=0.23$ (red), and  $k_pr_D=0.022$ (blue).}
\label{Solution_B}
\end{figure}

Figure~\ref{Solution_B} shows the magnetic potential as a function of $r$ at different  Debye radii. Note that $e\psi_0/T=1$ at $k_pr_D=1$, $e\psi_0/T\approx 4.3$ at $k_pr_D=0.23$, and $e\psi_/T\approx 11.8$ at $k_pr_D=0.022$. 
Analysis of the monotonic solution $\psi(r)$ at even smaller Debye radii suggests that  the magnetic potential at the filament center grows    approximately as 
\begin{eqnarray}  \label{eq:asympt_shooting}
e\psi_0/T\approx -4\ln k_pr_D-3.7,
\end{eqnarray}
 when $k_pr_D\rightarrow 0$. Therefore, near the filament center (at the distances $r\sim r_D$) the plasma return current  is much smaller than the beam current  ($e\psi_0/T<< 1/k_p^2r_D^2$). On the other hand, because the beam current  decreases faster than   the plasma return current, the latter becomes dominant at  distances $r>>r_D$. At even larger distances $r\gtrsim k_p^{-1}$, these currents reaching the minimum value become  equal to each other.

  This  currents' behavior   allows us to  find the approximate analytical solution of Eq.~(\ref{eq:axialEquilibrium}) at small $k_pr_D$.
 Omitting the first term in the
right-hand side of Eq.~(\ref{eq:Equilibrium}) near the filament center, we obtain 
\begin{eqnarray}  \label{eq:center_Equilibrium}
\frac{1}{ k_p^{2}r}\frac{d}{d r} r\Big(\frac{e\psi}{T}%
\Big)= - \frac{1}{k_p^2r_{D}^2}\exp\Big(\frac{e\psi}{T}-\frac{e\psi_0}{T}%
\Big).
\end{eqnarray}
This equation describes the Bennett distribution of beam particles with
uniform transverse beam temperature \cite{DavBook}: 
\begin{eqnarray}
e\psi/T=e\psi_0/T-2 \ln[( 1+(r/r_B)^2],  \label{eq:center_psi} \\
n_b=\frac{8/(k_pr_B)^2}{(1+( r/r_B)^2)^2}\frac{T}{mc^2}n_p,
\label{eq:center_nb}
\end{eqnarray}
where $r_B=2\sqrt{2}r_D$ is the radius of the Bennett pinch
The Bennett distribution is accurate at small distances from the filament
center, $r\sim r_B$, where magnetic screening is negligible. 

In the  range of distances $r_B\lesssim r \lesssim k_p^{-1}$, where the plasma
return current is dominant,
one can omit the second term in the right-hand
side of Eq.~(\ref{eq:Equilibrium}), 
\begin{eqnarray}  \label{eq:periphery_Equilibrium}
\frac{1}{ k_p^{2}r}\frac{d}{d r} r\frac{d}{d r} \Big(\frac{e\psi}{T}\Big)=\frac{e\psi}{T},\quad
\frac{e\psi}{T}=a K_0(k_p r),
\end{eqnarray}
where $a$ is some constant, and $K_0$ is the modified Bessel function of the
second kind. Solutions (\ref{eq:center_psi}) and (\ref%
{eq:periphery_Equilibrium}) should match at
distances, $r_B<<r<<k_p^{-1}$. The magnetic potential in the Bennett pinch has
the following asymptotics at $r>>r_B$: 
\begin{eqnarray}  \label{eq:asymptotic_center}
\frac{e \psi}{T}\approx \frac{e \psi_0}{T} +4\ln(k_pr_B) -4\ln k_pr,
\end{eqnarray}
while the solution (\ref{eq:periphery_Equilibrium}) at $r<<k_p^{-1}$ can be
approximated by: 
\begin{eqnarray}  \label{eq:asymptotic_periphery}
\frac{e\psi}{T}\approx a(-\gamma_E + \ln2 - \ln k_pr),
\end{eqnarray}
where $\gamma_E=0.5772...$ is the Euler constant. Comparison of (\ref%
{eq:asymptotic_center}) and (\ref{eq:asymptotic_periphery}) gives $a=4$
and the expression for the magnetic potential at the filament center,  
\begin{eqnarray}
\frac{e \psi_0}{T}=-4\ln(k_p r_B)+4(-\gamma_E + \ln2),  \label{eq:psi0}
\end{eqnarray}
which, after substitution $r_B^2=8r_D^2$, reduces to the expression Eq.~(\ref{eq:asympt_shooting}) found by the shooting method.
Using Eq.(\ref{eq:psi0}), we find that the  constant $n_{b*}$ in Boltzmann distribution at small $k_pr_D$ is given by 
\begin{eqnarray}
\frac{{n}_{b*}}{n_p}=\frac{T}{mc^2}\frac{\exp{(-e\psi_0/T)}}{k_p^2r_D^2}=\alpha_1{k_p^2r_B^2}\frac{T}{mc^2},
\label{eq:nbfaraway}
\end{eqnarray}
where $\alpha_1\equiv \exp(4\gamma_E)/2\approx 5.0$. At  large distances from the filament center where the beam and plasma currents become equal to each other, the magnetic potential $\psi_{\infty}$  is determined by the equation $e\psi_{\infty}/T=(mc^2/T)(n_{b*}/n_p)\exp{(e\psi_{\infty}/T)}$ and equal to 
\begin{eqnarray}
\frac{e\psi_{\infty}}{T}=-W(-\alpha_1 k_p^2r_D^2)\approx \alpha_1 k_p^2r_D^2,
\label{eq:PSI_faraway}
\end{eqnarray}
 where $W(x)$ is the Lambert W-function.
It is convenient to use the
interpolation formulas for the magnetic potential and beam particles distribution  which work at all distances and can be used even for moderately small
parameter $k_{p}r_{B}$,
\begin{eqnarray}
\frac{e\psi}{T} =4K_{0}[k_{p}(r^{2}+r_{B}^{2})^{1/2}]+\frac{e\psi_{\infty}}{T},\label{eq:interpPSI}\\
n_b=\frac{8/(k_pr_B)^2}{(1+( r/r_B)^2)^2}\frac{T}{mc^2}n_p +n_{b\infty}, \label{eq:interp_nb}
\end{eqnarray}
where $n_{b\infty}=(n_pT/{mc^2}){e\psi_{\infty}}/T$. These formulas  describe the modified Bennett distribution of the beam
particles in the presence of plasma return current.

\begin{figure}[t]
\includegraphics[height=0.14\textheight,width=.45\columnwidth]{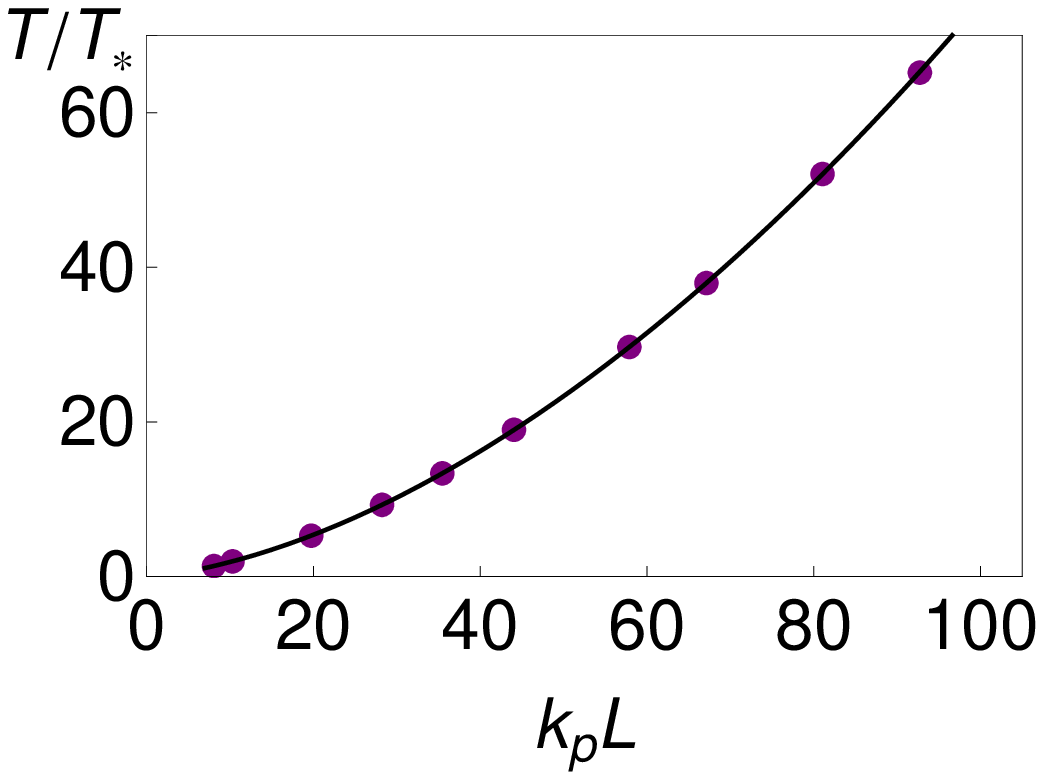}%
\hspace{0.1in},\includegraphics[height=0.14\textheight,width=.45\columnwidth]{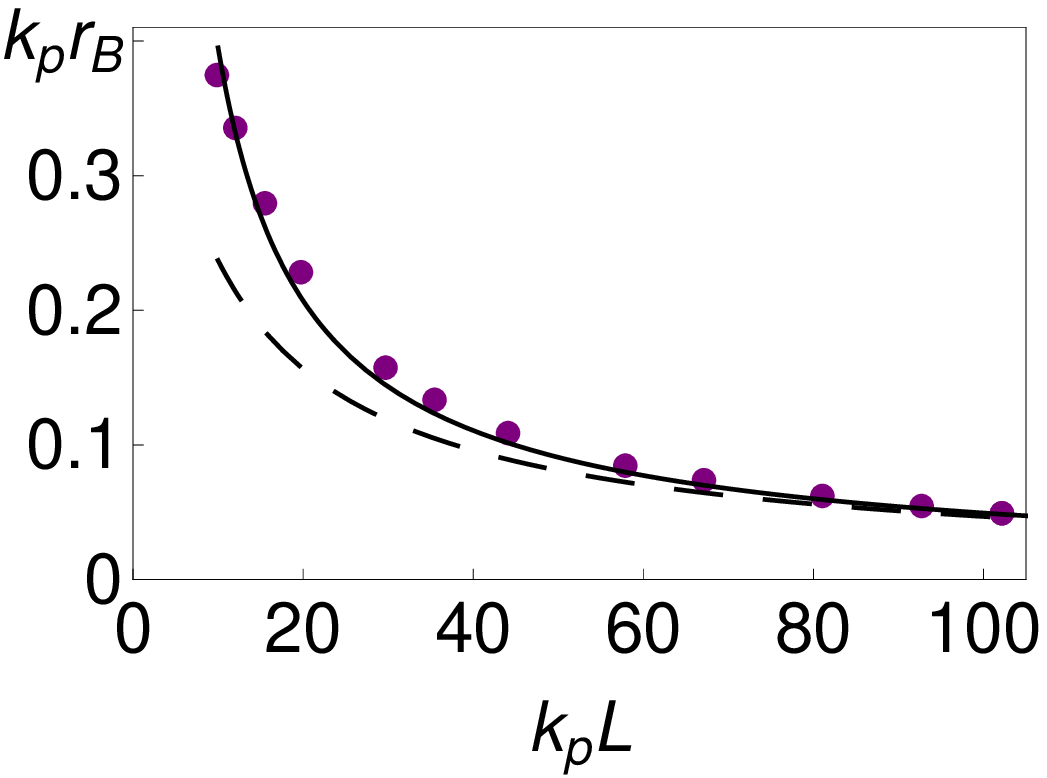} %
\caption{ The parameters $T$ and $r_B$  of the modified Bennett pinch as a function of the transverse beam dimension $L$. The pinch is obtained from Eq.~(\ref{eq:Equilibrium})   by the shooting method  (purple bullets) and analytically  (black lines). The dashed black line corresponds to the analytical formula for the radius without adjustment.}  
\label{fig:T_and_rB_on_L}
\end{figure}

We have found a general solution of Eq.(\ref{eq:Equilibrium}) which depends  on the transverse temperature of beam particles $T$, and the radius of the Bennett pinch $r_B$ (or $r_D$). 
 For an arbitrary transverse beam dimension, these two parameters can be determined from two conservation laws.  The procedure is straightforward when the solution of Eq.(\ref{eq:Equilibrium}) is found numerically; the result for is presented in Fig.~\ref{fig:comparison_density} by purple bullets.   To find analytical formulas, we introduce dimensionless variables  $\tilde\psi=e\psi/T$, $\tilde n=mc^2n_b/(T n_p)$, and dimensionless parameters $\tilde T=T/T_*$ (where $T_*=mc^2 n_{b0}/n_p$), $\tilde n_{\infty}=mc^2n_{b\infty}/(T n_p)$, and $\tilde \psi_0=e\psi_0/T$.  Substituting Eq.~(\ref{eq:interp_nb}) into  Eq.~(\ref{eq:particle_con_Max}), we can evaluate the number of particles as
\begin{eqnarray}
8\pi \tilde T +\tilde n_{b\infty}\tilde T k_p^2L^2=k_p^2L^2,\label{eq:N_asymptote}
\end{eqnarray}%
The first term in the left-hand side of  this equation is proportional to the number of the particle in the Bennett pinch and the second term is proportional to the number of particles outside of the pinch.   To evaluate the potential energy  at small $k_p r_B$, we use Eqs.~(\ref{eq:center_psi}) and (\ref{eq:center_psi})  and transform  Eq.~(\ref{eq:energy_con_Max}) to
\begin{eqnarray}
\tilde T k_p^2L^2- \frac{1}{2}\tilde T^2 (8\pi\tilde \psi_0-16\pi) =-\frac{1}{2}k_p^2L^2,\label{eq:W_asymptote}
\end{eqnarray}%
We can make further simplification, noting that $\tilde T>>1$ when $k_pL>>1$ and neglecting  the term in the right-hand side of Eq.(\ref{eq:W_asymptote}) proportional to the initial potential energy. Using Eq.~(\ref{eq:psi0})  and formula $\tilde n_{b\infty}=\alpha_1 k_p^2r_B^2$, we find the temperature and the radius
\begin{eqnarray}
{T}=  \frac{k_p^2L^2}{1+W(\alpha_0 {k_p^2L^2})}\frac{T_*}{8\pi},\label{eq:T_on_L}\\
k_p^2r_B^2=\frac{8\pi}{\alpha_1 k_p^2 L^2}W[\alpha_0 (k_pL+A_1)^2]. \label{eq:rB_on_L}
\end{eqnarray}%
Surprisingly despite all approximations, formula (\ref{eq:T_on_L})  very closely reproduces the result of the numerical integration of Eq.~(\ref{eq:Equilibrium}), see Fig.~\ref{fig:T_and_rB_on_L} (left panel). To improve accuracy of the other formula, we introduce   the adjustment parameter $A_1$ into Eq.~(\ref{eq:rB_on_L}), otherwise it is not accurate at moderately large $k_pL$. Comparison of the adjusted ($A_1=35$) and unadjusted ($A_1=0$) formulas with the result obtained by the shooting method is given  on the right panel in this figure. 

Now we find several integral characteristics of the modified Bennett pinch. It is interesting to note that the total magnetic energy $W_m$ in this  pinch is a finite quantity (in contrast to the regular Bennett pinch, where it diverges at large distances).  Using Eq.~(\ref{eq:interpPSI}) we calculate $\vec{B}=-\nabla\psi$ and then integrate $B^2/8\pi$ over the transverse area. The result can be expressed in special functions; with some adjustments,  it is given at   at  small  $k_pr_B$ by
\begin{eqnarray}
W_m=\frac{T^2}{4\pi e^2}(-44.4 +66 k_p r_B-16\pi\ln k_pr_B). \label{eq:mag_rB}
\end{eqnarray}   
The number of particles trapped in the modified Bennett pinch can be found by integrating  Eq.~(\ref{eq:interp_nb}) over the transverse area. Using Eq.~(\ref{eq:T_on_L}) and making some adjustments,  we find   
\begin{eqnarray}
N_{tr}=N\frac{1+0.83 r_B}{1+W(\alpha_0 k_p^2L^2)}. \label{eq:number_Ntr_rB}
\end{eqnarray} 
where $r_B$ should be calculated from  Eq.(\ref{eq:rB_on_L}). Obviously, the number of untrapped particles  is given by
\begin{eqnarray}
N_{untr}=N\frac{W(\alpha_0 k_p^2L^2)-0.83 r_B}{1+W(\alpha_0 k_p^2L^2)}. \label{eq:number_Nuntr_rB}
\end{eqnarray}



\bigskip

\begin{acknowledgements}
This work was supported by the US DOE grant DE-FG02-05ER54840. We thank E.
Startsev for fruitful discussions.
\end{acknowledgements}

\end{document}